%% file: TOP-11-004_temp.tex
\pdfoutput=1

\documentclass[11pt,twoside,a4paper,cmspaper,final,collab]{cms-tdr}
\def\svnVersion{185700}\def\cmsCernNo{2012-374}\def\cmsCernDate{\today}\def\cmsMessage{Submitted to the European Physical Journal C}

\begin{document}\cmsNoteHeader{TOP-11-004}

\hyphenation{had-ron-i-za-tion}
\hyphenation{cal-or-i-me-ter}
\hyphenation{de-vices}

\RCS$Revision: 176925 $
\RCS$HeadURL: svn+ssh://svn.cern.ch/reps/tdr2/papers/TOP-11-004/trunk/TOP-11-004.tex $
\RCS$Id: TOP-11-004.tex 176925 2013-03-17 16:49:21Z alverson $
\providecommand{\tauh}{\ensuremath{\Pgt_\mathrm{h}}\xspace}
\newlength\cmsFigWidth
\ifthenelse{\boolean{cms@external}}{\setlength\cmsFigWidth{0.95\columnwidth}}{\setlength\cmsFigWidth{0.7\linewidth}}
\cmsNoteHeader{TOP-11-004} 
\title{Measurement of the \ttbar production cross section in the $\tau$+jets channel in pp collisions at $\sqrt{s}=7$\TeV}

\date{\today}

\abstract{
The top-quark pair production cross section in 7\TeV centre-of-mass energy proton proton collisions is measured using data collected by the CMS detector at the LHC.
The measurement uses events with one jet identified as a hadronically-decaying $\tau$ lepton
and at least four additional energetic jets, at least one of which is identified as coming from a
b quark. The analysed data sample corresponds to an
integrated luminosity of 3.9\fbinv recorded by a dedicated multijet plus hadronically-decaying $\tau$ trigger.
A neural network has been developed to separate the top-quark pairs from the W+jets and multijet backgrounds.
The measured value of $\sigma_{\ttbar} = 152\pm12\stat\pm
32\syst\pm3\lum\unit{pb}$ is consistent with the standard model predictions.
}

\hypersetup{%
pdfauthor={CMS Collaboration},%
pdftitle={Measurement of the top-antitop production cross section in the tau+jets channel in pp collisions at sqrt(s) = 7 TeV},%
pdfsubject={CMS},%
pdfkeywords={CMS, physics, top pair production}}

\maketitle 

\section{Introduction}

Top-quark pairs (\ttbar) are copiously produced at the Large Hadron Collider (LHC) primarily through gluon-gluon fusion.
The measurements of the \ttbar production cross section and branching fractions are important tests of the standard model (SM),
since the top quark is expected to play a special role in various extensions of the SM due to its high
mass (see for example~\cite{Agashe:2006hk,Hill:1993hs}). 

The branching fraction of a top-quark decay to a W boson and a b quark is close to 100\% in the SM.
Therefore, the final states from the top-quark decays are given by the decay mode of the W bosons.
In this Letter top-quark decays in the ``hadronic $\tau+\text{jets}$" final state are studied.
One W boson decays into a hadronically-decaying $\tau$ lepton ($\tauh$) and a
neutrino with a branching fraction of $0.1125\times0.647$~\cite{PDG} and the other one decays to a quark-antiquark pair with a branching
fraction of 0.676~\cite{PDG}. Thus, 9.8\% of the \ttbar pairs produced are expected to lead to this final state.

The branching fraction of \ttbar to $\tauh$+jets final states is expected to be the largest one among
those with $\tau$ leptons in the final state.
The existence of charged Higgs bosons could give rise to an enhanced cross section in this
channel. The top quark would decay via $\cPqt \to\PH^+ \cPqb$ and the charged Higgs boson subsequently decays to a $\tau$ lepton via $\PH^+\to\tau^+\nu_{\tau}$.
The present status of the charged Higgs boson search in \ttbar final states with the Compact Muon Solenoid (CMS) detector is described in Ref.~\cite{HIG-11-019}.

In this Letter we present a measurement of the
\ttbar production cross section in the $\tauh$+jets final state in
proton-proton collisions at $\sqrt{s}=7$\TeV using data collected by the CMS experiment.
It is the first such measurement performed using the CMS detector and complements the measurement performed in the $\tauh+\text{lepton}$ channel~\cite{TOP-11-006}.
The \ttbar production cross section in the $\tauh$+jets final state has previously been measured in proton-antiproton collisions at
$\sqrt{s}=1.96$\TeV at the Tevatron~\cite{D0,CDF} and more recently in proton-proton collisions at $\sqrt{s}=7$\TeV using the ATLAS
detector~\cite{ATLAS}. All measurements referenced above have been found to be in agreement with the SM expectations.

\section{The CMS detector}
\label{sec_CMS}

The central feature of the CMS apparatus is a superconducting cylindrical solenoid 6 m in diameter, which provides an axial magnetic field of
3.8\unit{T}. Within the field volume are the silicon pixel and strip trackers, the crystal electromagnetic (ECAL) and brass/scintillator hadronic (HCAL) calorimeters which provide identification of charged, electromagnetic and hadronic particles up to pseudorapidities of $|\eta|<2.5$
(trackers) and $|\eta|<3.0$ (calorimeters).
The pseudorapidity is defined as $\eta=-\ln[\tan (\theta /2)]$, where $\theta$ is the polar angle measured with respect
to the positive $z$ axis of the right-handed coordinate system used by the CMS
experiment. The $x$ axis
points towards the centre of the LHC ring, the $y$ axis is directed upward along the vertical and
the $z$ axis corresponds to the anticlockwise-beam direction.
 In addition the CMS detector has extensive forward calorimetry.
Muons are measured in gas detectors embedded in the steel return yoke outside the solenoid. The excellent tracker impact
parameter resolution of $\approx$15\mum and transverse
momentum (\pt) resolution of $\approx$1.5\% for 100\GeV particles support a robust identification of
$\tauh$ and jets arising from b quark hadronisation. A detailed description of the CMS detector can be found elsewhere~\cite{detpaper}.

\section{Event simulation}
\label{sec_evt_simu}

Monte Carlo (MC) simulation is used to determine the signal efficiency as well as the contribution from electroweak and \ttbar
background processes (\ie contributions from the full hadronic, $\text{lepton}+\text{jets}$,
$\tauh+\text{lepton}$ and $\tauh\tauh$ channels).
The \ttbar signal and background events as well as the $\PW/\cPZ+\text{jets}$ events are simulated using the \MADGRAPH (v.5.1.1.0)~\cite{madgraph}
generator using the CTEQ6L1~\cite{cteq} parton distribution functions (PDFs).
The simulation of parton showering, fragmentation, hadronisation
and decays of short-lived particles, except $\tau$ leptons, is performed by \PYTHIA
(v.6.424)~\cite{pythia}. Tau lepton decays are simulated using \TAUOLA (v.2.75)~\cite{tauola}.
Single top-quark events are generated using \POWHEG (r1380)~\cite{powheg} interfaced to \PYTHIA and \TAUOLA. The top-quark mass is
set to 172.5\GeV, and the approximate next-to-next-to-leading-order (NNLO) \ttbar production cross section of $164 \pm 10$\unit{pb} is calculated using the MSTW2008
next-to-next-leading-log PDFs~\cite{ttbartheo}.
Simulated events are weighted to reflect the distribution of the number of multiple interactions (pileup) observed in data.
Data-to-simulation b-tagging efficiency scale factors are applied to correct for the differences between data and simulation.
\section{Dataset}
\label{sec_data_set}

The total integrated luminosity of the dataset analysed is 3.9\fbinv.
A multijet trigger, in which one of the jets is required to be identified as a hadronically-decaying $\tau$ lepton, was designed to record
$\Pp\Pp\to\ttbar\rightarrow\tauh+\text{jets}$ events.
It consists of two consecutively applied filters, referred to as jet and $\tau$ filters.
The jet filter requires the presence of four central jets reconstructed in the calorimeter ($|\eta|<2.5$, $\pt > 40\GeV$),
referred to as calorimeter jets.
The $\tau$ filter requires the presence of one isolated
particle-flow~\cite{PAS-PFT-09-001} $\tau$ candidate ($|\eta|<2.5$, \pt $>$ 40\GeV, at least one track
with $\pt > 5\GeV$), matched to one of four trigger jets.
Due to the increasing rate of the recorded events with
the rising instantaneous luminosity, the thresholds on the jets and $\tau$
lepton were raised to $\pt > 45\GeV$ during the later part of the data taking period. About 80\% of the data were recorded with
that more restrictive trigger configuration.
The overall $\Pp\Pp\to\ttbar\rightarrow\tauh+\text{jets}$ trigger efficiency is small, approximately
1\%, with respect to all generated $\ttbar\to\tauh+\text{jets}$ events.
The small efficiency is due to the high \pt threshold on the hadronically-decaying $\tau$ lepton.

\section{Event selection \label{sec_evt_sel}}

The object reconstruction relies on the particle-flow technique.
The event selection is based on the presence of at least four particle-flow jets, reconstructed with the anti-\kt
clustering algorithm~\cite{antikT1, antikT2} with a distance parameter of $R=0.5$, and on the presence of one particle-flow $\tau$
candidate reconstructed with
the hadron-plus-strip (HPS)
identification algorithm~\cite{PAS-TAU-11-001}. The HPS algorithm exploits the ability of the particle-flow to reconstruct resonances in the $\tau$ decay.
It considers candidates with one or three charged hadrons and up to two neutral pions,
with a net charge of ${\pm}1e$.

The $\tau$ candidates are required to be isolated: the sum of the transverse energies of the additional charged hadrons and photons
($\tau$ decay products
excluded) reconstructed in an isolation cone of $\Delta R=\sqrt{(\Delta\eta)^2+(\Delta\phi)^2}=0.5$ (where $\phi$ is the azimuthal angle in
radians) around the $\tau$ candidate should be less than 1\GeV.
The $\tau$ reconstruction efficiency is estimated to be approximately 44\% for $\tauh$ candidates
 with \pt$>$ 20\GeV, $|\eta|<$2.3, selected in genuine $\cPZ\to\tau^+\tau^-$ events, with a corresponding misidentification efficiency for jets
 of 0.5\%~\cite{PAS-TAU-11-001}.
 
Furthermore, $\tau$ candidates are required to pass discriminators against muons and electrons.
The discrimination against electrons relies on a boosted decision tree
that combines variables that characterise the presence of neutral
particles reconstructed in the $\tau$ decay (\eg, number of constituents, cluster shapes, energy fractions), as well as the presence of a charged hadron and
electromagnetic particles (\eg, energy fractions, electron-pion discriminator).
To suppress the contamination from muons, the leading track of the $\tauh$ candidate is vetoed if identified as a muon in the muon detectors.
In addition, a single-charged $\tauh$ candidate should not be identified as a minimum ionising particle: the ratio of the sum of the energy deposits in the ECAL and HCAL calorimeters associated to the $\tauh$ candidate over the leading
track momentum of the $\tauh$ candidate should be larger than 0.2.

Three jets are required to have $\pt> 45\GeV$, $|\eta|<2.4$ and the
$\tauh$ candidate $\pt> 45\GeV$, $|\eta|<2.3$.
The offline jets and offline $\tauh$ candidates are explicitly matched to the jets and $\tauh$ candidates used by the trigger. The presence of an additional jet with $\pt > 20\GeV$ ($\tauh$ candidate
excluded) is required.

Since two b-jets from the top-quark decays are expected in the final state, at least one jet is required to be identified as a b-jet using the medium working point of
the jet probability algorithm~\cite{BTV-11-004}. At a misidentification probability for
light-flavoured jets of 1\%, a b-tagging efficiency of 60\% is achieved for this working point of the tagging algorithm.

A veto on the presence of loosely isolated electrons and muons is applied to further prevent the misidentification of genuine electrons
and muons as $\tauh$ candidates.
The isolation requirement is defined as $I/\pt<0.15$, where $I$ is the sum of the transverse energy deposits in the ECAL and HCAL
calorimeters and \pt is the scalar value of the track momenta in a cone with $\Delta R=0.3$ centred on the lepton direction, excluding the lepton \pt.

The momentum imbalance, $\ptvec^{\text{miss}}$, is defined as the opposite of the
vectorial sum of the particle transverse momenta, using all particles
reconstructed by the particle-flow algorithm. The transverse
missing energy, \MET, is defined as the magnitude of this quantity
and is required to be greater than 20\GeV to reject the multijet background and
to achieve a good separation for the input variables used in the artificial neural network described in Section~\ref{sec_bkg}.
Events which pass this set of criteria constitute the preselected sample, from which the yield is extracted.

The trigger efficiencies have been measured in data,
determining separately the efficiency of a single jet and a single $\tauh$ to pass the trigger requirements.
The single-jet efficiency has been measured in events
containing four particle-flow jets in the central region, three of them matched to the trigger jets.
The fourth jet is used as a probe jet and the single-jet efficiency is computed with respect to its match to the fourth trigger jet.
The efficiency of a single particle-flow jet with $\pt\approx 45\GeV$ to pass the single-jet requirement of the trigger is $70\pm1$\%
($54\pm1$\% for the more restrictive trigger). The jet trigger plateau is reached above $\approx$120\GeV due to
the different energy scale of particle-flow jets and calorimeter jets.

The $\tauh$ trigger efficiency has been measured in the events that satisfy the jet filter requirement and
that contain a reconstructed $\tauh$ candidate matched to one of the four trigger jets.
The $\tauh$ trigger plateau is reached for $\pt >45\GeV$ (respectively $\pt > 50\GeV$ for the more stringent trigger)
yielding an efficiency of $90\pm1$\% ($92\pm1$\%). The trigger efficiency is modeled in simulation by multiplying the trigger efficiencies obtained for the three most energetic central jets and the trigger efficiency obtained for the $\tauh$
candidate.

\section{Background estimation \label{sec_bkg}}

The largest background for this analysis comes from high-multiplicity multijet events where one of the jets is
misidentified as a $\tauh$, and represents approximately 90\% of the expected background.
While control samples in data are used to evaluate the multijet background, the estimation of the other contributions from \ttbar backgrounds and electroweak
processes, such as
single top-quark production and $\PW/\cPZ+\text{jets}$ events, relies on MC simulation. Given the low expected signal over
background ratio expected after preselection, an artificial neural network (ANN) is used to discriminate signal
and background.

\subsection{Multijet background \label{sec_multijet} }

The multijet background is estimated from data by using the same
selection as the preselected sample except that a veto is applied on the presence of a b-tagged jet.
From simulated events, we expect the resulting sample, referred to as multijet sample, to contain less than 0.6\% of $\ttbar\rightarrow\tauh+\text{jets}$ events, less
than 0.3\% of \ttbar background events and less than 2.0\% of $\PW+\text{jets}$ and $\cPZ+\text{jets}$ events. Therefore the multijet sample provides a good
representation of the multijet background and is used to train the ANN.

To account for the kinematic bias of the b-tag veto in the multijet sample, as the b-tagging efficiency depends on the jet
momenta, the selected multijet events in data are weighted by the misidentification probability to
select at least one b-jet in the event. This assumes that the jets are predominantly light flavoured:
\ifthenelse{\boolean{cms@external}}
{
\begin{multline*}
P(\text{number of misidentified jets} \geq 1) =\\
\sum_{i=1}^{N_{\text{jet}}} [P_i \cdot
\Pi_{j}^{N_{\text{jet}}}(1-P_j)]
\end{multline*}
}
{
$P(\text{number of misidentified jets} \geq 1) =\sum_{i=1}^{N_{\text{jet}}} \left[P_i \cdot
\Pi_{j}^{N_{\text{jet}}}(1-P_j)\right]$
}
with $j \not= i$, where $P_j$ stands for the
misidentification probability of a light-flavoured jet and has been
measured for different \pt and $\eta$ bins in control samples in data~\cite{BTV-11-004}.

\subsection{Artificial neural network \label{sec_NN}}

The following seven variables are used to build an artificial neural network: the scalar sum of the transverse momenta of all the
selected jets and the $\tauh$, \HT,
the aplanarity,
 the $\tauh$ charge multiplied by the absolute value of the pseudorapidity of the $\tauh$ candidate,
 $q(\tauh)\cdot|\eta(\tauh)|$, the missing transverse energy,
\MET, the azimuthal angle between the $\tauh$ candidate and the missing transverse energy direction,
$\Delta\phi(\tauh,\ptvec^{\text{miss}})$, the invariant mass of the system of all the selected jets and the $\tauh$ candidate,
$M(\text{jets},\tauh)$, and the $\chi^2$ returned by a kinematic fit constraining the
hadronically-decaying $\PW$ boson and top-quark masses to $m_\PW=80.4$ \GeV and $m_{\text{top}}=172.5$ \GeV.
The aplanarity, $A=\frac{3}{2}\lambda_1$, is used to describe the spherical topology of the top-quark decay products:
$\lambda_1$ is the smallest eigenvalue of the momentum tensor $M^{\alpha\beta}=\sum_{i}p^{\alpha}_{i}p^{\beta}_{i}/\sum_i|\vec{p_{i}}|^2$, where $i$ runs over the number of jets and the
$\tauh$ candidate and $\alpha,\beta=1,2,3$ specify the three spatial components of the momentum.
The $\tauh$ charge multiplied by the absolute value of the pseudorapidity of the $\tauh$ candidate, $q(\tauh)\cdot|\eta(\tauh)|$, is used to account for the charge-symmetric nature of \ttbar events in contrast to $\PW+\text{jets}$ events produced in
 proton-proton collisions. The $\tauh$ charge is defined as the sum of the charges of the charged hadrons selected by the HPS algorithm.
 The training is performed using simulated $\ttbar\to \tauh+\text{jets}$ events passing the preselected sample criteria and events from the
multijet sample.

\subsection{Signal yield extraction\label{sec_fit} }

To minimise the statistical uncertainty of the cross section measurement, we fit the entire ANN output, $D_{\mathrm{NN}}$,
distribution rather than counting events above a given value.
The extraction of the yield is performed via a two-component  binned negative log-likelihood fit of the shapes of expected signal
and multijet background, derived respectively from
simulation and the multijet sample, to the data. The
shapes for the \ttbar background and the electroweak processes, and their
normalizations are fixed to the expectation from simulation.
Table~\ref{tab_fit} summarizes the contribution of the various processes. The number of signal
 events among the 3050 selected events is $383 \pm 29$.
The fit uncertainty is given for the number of signal and multijet events, whereas for the
remaining backgrounds the statistical uncertainty is due to the limited size of the simulated
samples. The systematic uncertainties correspond to those described in Section~\ref{sec_uncertainties}.

\begin{table}[!h]
\begin{center}
\topcaption{Estimated number of signal and multijet events after a fit to the ANN output distribution, and expected contributions of the electroweak processes and
\ttbar backgrounds from MC simulation.}
\label{tab_fit}
\begin{tabular}{l l  }
\hline\noalign{\smallskip}
Source & Events \\
\noalign{\smallskip}\hline\noalign{\smallskip}
Signal $\ttbar\to\tauh+\text{jets}$ & $383\pm 29\,\text{(fit)}$\\
\noalign{\smallskip}\hline\noalign{\smallskip}
Multijet & $2392\pm 29\,\text{(fit)}\pm120\syst$ \\
Other \ttbar & $151\pm4\stat\pm37\syst$ \\
W+jets & $62\pm8\stat\pm14\syst$ \\
Single top & $41\pm1\stat\pm8\syst$ \\
Z+jets & $21\pm2\stat\pm4\syst$ \\
\noalign{\smallskip}\hline\noalign{\smallskip}
Total backgrounds &  $2667\pm31\stat \pm 127\syst$ \\
\noalign{\smallskip}\hline\noalign{\smallskip}
Data & 3050 \\
\noalign{\smallskip}\hline
\end{tabular}
\end{center}
\end{table}

\begin{figure}[htb]
\centering\includegraphics[width=\cmsFigWidth]{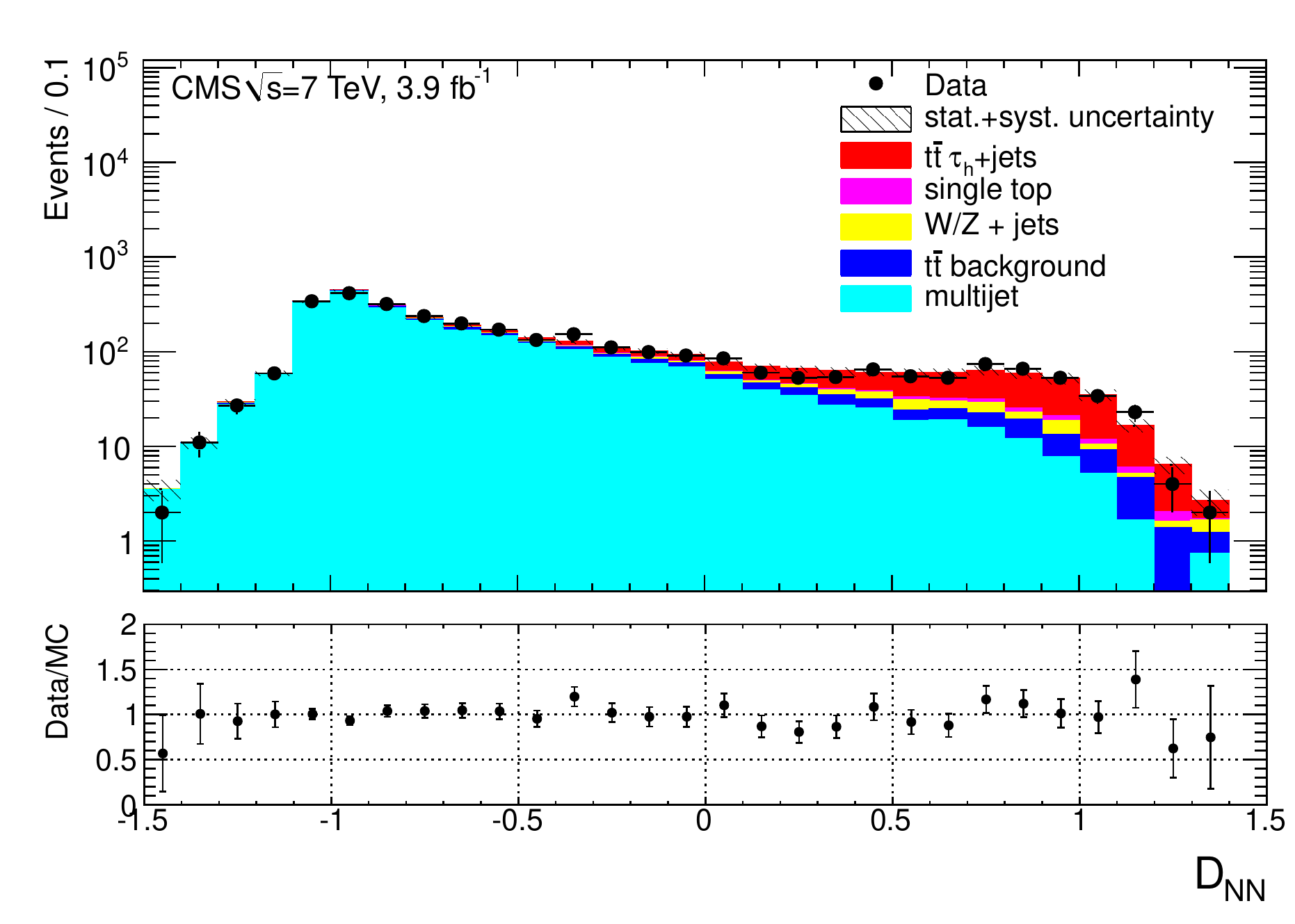}
\caption{Distribution of the artificial neural network output variable after a fit of the signal and multijet processes to the
data. Other background shapes and normalisations are fixed to the expectations from simulation.
The hatched area shows the combined statistical and systematic uncertainty on the sum of the signal and background predictions. The
ratio of the data distribution to the sum of expected background and fitted signal distributions is shown at the bottom of the Figure.}
\label{fig_NN}
\end{figure}

\begin{figure}[htb]
\centering\includegraphics[width=\cmsFigWidth]{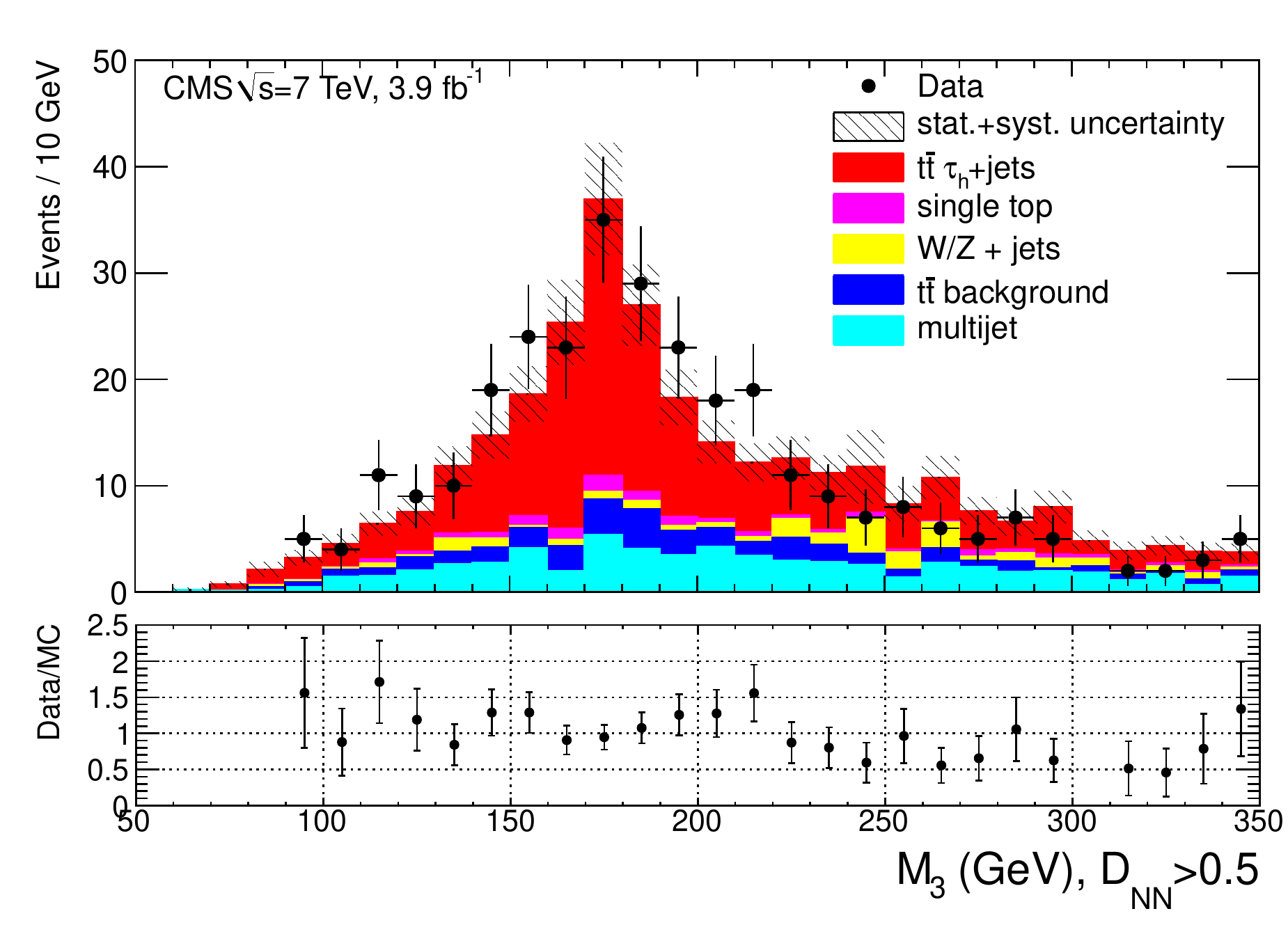}

\caption{Distribution of the reconstructed $M_3$ variable after a fit of the signal and multijet processes to the data, after requiring the ANN
output value to be greater than 0.5. The hatched area shows the combined statistical and systematic uncertainty on the sum of the signal and background predictions.
The ratio of the data distribution to the sum of expected background and fitted signal distributions is shown at the bottom of the Figure.}
\label{fig_njets}
\end{figure}

Figure~\ref{fig_NN} shows the fitted ANN output distribution.
Figure~\ref{fig_njets} shows the distribution of $M_3$, defined as the invariant mass of the three-jet system with highest transverse
momentum in an enriched signal region, $D_{\mathrm{NN}}>0.5$. The selected jets
are deemed to originate from the hadronically-decaying top quark.

\section{Cross section measurement \label{sec_xsec}}
\subsection{Systematic uncertainties \label{sec_uncertainties}}

The main sources of systematic uncertainties are those due to uncertainty in the jet energy
scale (JES), the $\tauh$ energy scale, the $\tauh$ identification, the trigger efficiency
and in the \MET measurement.
The uncertainty in the cross section measurement is obtained by
 combining the uncertainty in the signal acceptance and in the fitted number of signal events.
The systematic uncertainties in the fitted number of signal events are estimated, when relevant, by iterating the fit on the ANN output
in order to take into account possible shape variations of the ANN input variables.

The uncertainties in the cross sections for the
  different simulated background processes are estimated from theoretical calculations~\cite{ttbartheo, Gavin:2010az}.
The uncertainty coming from the top-quark mass is evaluated considering two simulated samples where the nominal
top-quark mass of 172.5\GeV has been shifted by $\pm$6\GeV.
Scaling this uncertainty to the measured top-quark mass uncertainty of 1.1\GeV provides a 3\% relative uncertainty in the measured cross section.
The dependence of the selection on the renormalization and factorization scales is estimated by varying these scales
simultaneously by a factor of 0.5 and 2.0 from their default value equal to the hard-scattering scale $Q$, with $Q^2=m^2_{\text{top}}+\sum\pt^2$, where $m_{\text{top}}$ denotes the
top-quark mass and $\sum\pt^2$ the sum of the squared transverse momenta of all final state partons.
The measured relative uncertainty for the \ttbar processes is estimated to be 2\%.

The thresholds used to associate the matrix elements to the parton showers are varied
from 20\GeV to respectively 10\GeV and 40\GeV. The measured relative uncertainty for the \ttbar processes is estimated to be
3\%. The uncertainty of the choice of PDFs on the signal acceptance is estimated using the $2\times11$  reference PDFs associated to CTEQ6L1.
The uncertainty of the choice of PDFs on the number of fitted signal
events is determined iterating the fit on the ANN output
distribution.  Simulated events using the reference PDFs (out of the
eleven available) leading to the maximal up (respectively maximal down) variation are used.

The uncertainty induced by the statistical uncertainty of the trigger turn-on is computed using the uncertainties
on the trigger turn-on curves versus the transverse momenta of the particle-flow jets and particle-flow $\tauh$.
An additional $\pm$5\% uncertainty is assigned to the $\tauh$ trigger efficiency measurement,
since the data used to estimate the $\tau$-leg efficiency consist mainly of jets misidentified as $\tauh$ candidates.
This uncertainty is derived in comparison to the trigger efficiency obtained for $\tau_h$ candidates in genuine $\cPZ\to\tau^+\tau^-$ events using
similar trigger conditions.

The pileup uncertainty is estimated by varying the number of pileup interactions measured in data
according to the theoretical uncertainty of the minimum bias inelastic cross section of $\pm$8\%.

The uncertainty on the $\tauh$ energy scale is estimated by varying the $\tauh$ energy by $\pm$3\%~\cite{PAS-TAU-11-001}. The
uncertainties are propagated to the trigger efficiency weights. The uncertainty due to the $\tauh$ identification efficiency is estimated to be 6\%~\cite{PAS-TAU-11-001}.

The uncertainty due to the JES is estimated by rescaling the jet energy up or down by the uncertainties corresponding to one standard deviation.
For the jet energy resolution (JER) the distribution of the jet energy has been smeared by one standard deviation. The corrections are propagated to the \MET measurement and to the trigger-efficiency measurement.
The energy of the particles that are not clustered into jets is varied by $\pm$10\%, leading to an additional uncertainty in the \MET.

The uncertainty due to applying b-tagging data-to-simulation scale factors for b, c and
light-flavoured jets to the simulated events is estimated
  by shifting the value of the applied scale factors by the uncertainty corresponding to one standard deviation~\cite{BTV-11-003}.
  The uncertainty in the reweighting method applied to the multijet data sample is estimated to be 5\%.

The uncertainty in the luminosity measurement is estimated to be 2.2\%~\cite{lumi}.

Table~\ref{sys_unc} summarizes the uncertainties entering the cross section measurement, split into systematic and
statistical ones. The statistical uncertainty includes the $D_{\mathrm{NN}}$ fit uncertainty, the statistical uncertainty of the trigger turn-ons, as well as the uncertainty due to the limited size of the
simulated samples.

\begin{table}[tbh]
\centering
\topcaption{Relative uncertainties in the cross section measurement.\label{tab:sys_A}}
\label{sys_unc}
\scalebox{0.95}{
\begin{tabular}{lc}
\hline\noalign{\smallskip}
Source & Rel. uncert. [\%] \\
\noalign{\smallskip}\hline\noalign{\smallskip}
W/Z/\ttbar backgr. cross section uncert.& $\pm$3 \\
Top-quark mass         & $\pm$3     \\
Renormalization/factorization scale            & $\pm$2     \\
Parton matching        & $\pm$3     \\
PDF                    & $\pm$5    \\
$\tauh$ trigger efficiency        & $\pm$7    \\
Pileup                 & $+$5 $-$1    \\
$\tauh$ energy scale   & $\pm$7   \\
$\tauh$ identification & $\pm$9    \\
Jet energy scale       & $\pm$11   \\
Jet energy resolution  & $\pm$2    \\
Unclustered \MET       & $\pm$7    \\
b-tagging              & $\pm$3    \\
Multijet background reweighting & $\pm$5\\
\noalign{\smallskip}\hline\noalign{\smallskip}
Syst. uncert.            & $\pm$21 \\
\noalign{\smallskip}\hline\noalign{\smallskip}
Stat. uncert. from fit and MC samples & $\pm$8    \\
Stat. uncert. from trigger     & $\pm$0.4   \\
\noalign{\smallskip}\hline\noalign{\smallskip}
Total stat. uncert. & $\pm$8               \\
\noalign{\smallskip}\hline
\end{tabular}
}
\end{table}

\subsection{Measured cross section and branching fraction}

The measurement of the \ttbar cross section in the $\tauh$+jets channel is performed using the following expression:
$$\sigma_{\ttbar}=\frac{N-N_B}{A_{\text{tot}}\cdot \mathcal{B} \cdot \int \mathcal{L}\,\rd{}t}$$ where $N$ is the number of observed
candidate events, $N_B$ is the estimate of
the background, $\int\mathcal{L}\,\rd{}t$ is the integrated luminosity, $A_{\text{tot}}$
is the total acceptance, which contains the trigger efficiency and the efficiency of the offline event
selection and $\mathcal{B}$ is the branching fraction of the $\tauh$+jets channel.

Taking into account the systematic and statistical uncertainties reported in Table~\ref{tab:sys_A} and the evaluated acceptance,
$A_{\text{tot}} = 0.0066\pm0.0001\stat\pm0.0010\syst$, the cross section is:
$$\sigma_{\ttbar} = 152\pm12\stat\pm32\syst\pm3\lum\unit{pb}.$$

Using the number of fitted signal events and the theoretical \ttbar production cross section, the
branching fraction of the $\tauh+\text{jets}$ channel is:
$$\mathcal{B} = 0.091\pm0.007\stat\pm0.020\syst\pm0.002\lum.$$
The theoretical uncertainty on the \ttbar cross section is included in the systematic uncertainties.

\section{Summary}

Top-quark pairs in the $\tauh$+jets final state have been selected in a data sample from
 proton-proton collisions at $\sqrt{s}=7$\TeV, corresponding to an integrated luminosity of 3.9\fbinv. Events
 were recorded by a dedicated multijet plus $\tauh$ trigger, where events are selected with a
 moderate amount of \MET and four jets, at least one of which is b-tagged. The multijet background is discriminated against using an
 artificial neural network technique.
The result, $\sigma_{\ttbar} = 152\pm12\stat\pm32\syst\pm3\lum\unit{pb}$,  is consistent with CMS measurements performed in other \ttbar final
states~\cite{Chatrchyan:2011yy,Chatrchyan:2011ew,TOP-11-006,:2012bta}, as well as with the theoretical
NNLO value of $164\pm10\unit{pb}$.
The measured process is the dominant background to a charged Higgs search, where a
significant deviation from the SM expectations would indicate the presence of new phenomena.

\section*{Acknowledgements}
We congratulate our colleagues in the CERN accelerator departments for the excellent performance of the LHC and thank the technical and administrative staffs at CERN and at other CMS institutes for their contributions to the success of the CMS effort. In addition, we gratefully acknowledge the computing centres and personnel of the Worldwide LHC Computing Grid for delivering so effectively the computing infrastructure essential to our analyses. Finally, we acknowledge the enduring support for the construction and operation of the LHC and the CMS detector provided by the following funding agencies: BMWF and FWF (Austria); FNRS and FWO (Belgium); CNPq, CAPES, FAPERJ, and FAPESP (Brazil); MEYS (Bulgaria); CERN; CAS, MoST, and NSFC (China); COLCIENCIAS (Colombia); MSES (Croatia); RPF (Cyprus); MoER, SF0690030s09 and ERDF (Estonia); Academy of Finland, MEC, and HIP (Finland); CEA and CNRS/IN2P3 (France); BMBF, DFG, and HGF (Germany); GSRT (Greece); OTKA and NKTH (Hungary); DAE and DST (India); IPM (Iran); SFI (Ireland); INFN (Italy); NRF and WCU (Republic of Korea); LAS (Lithuania); CINVESTAV, CONACYT, SEP, and UASLP-FAI (Mexico); MSI (New Zealand); PAEC (Pakistan); MSHE and NSC (Poland); FCT (Portugal); JINR (Armenia, Belarus, Georgia, Ukraine, Uzbekistan); MON, RosAtom, RAS and RFBR (Russia); MSTD (Serbia); SEIDI and CPAN (Spain); Swiss Funding Agencies (Switzerland); NSC (Taipei); ThEPCenter, IPST and NSTDA (Thailand); TUBITAK and TAEK (Turkey); NASU (Ukraine); STFC (United Kingdom); DOE and NSF (USA).
\bibliography{auto_generated}   

\cleardoublepage \appendix\section{The CMS Collaboration \label{app:collab}}\begin{sloppypar}\hyphenpenalty=5000\widowpenalty=500\clubpenalty=5000\input{TOP-11-004-authorlist.tex}\end{sloppypar}
\end{document}

%% file: TOP-11-004-authorlist.tex
\textbf{Yerevan Physics Institute,  Yerevan,  Armenia}\\*[0pt]
S.~Chatrchyan, V.~Khachatryan, A.M.~Sirunyan, A.~Tumasyan
\vskip\cmsinstskip
\textbf{Institut f\"{u}r Hochenergiephysik der OeAW,  Wien,  Austria}\\*[0pt]
W.~Adam, E.~Aguilo, T.~Bergauer, M.~Dragicevic, J.~Er\"{o}, C.~Fabjan\cmsAuthorMark{1}, M.~Friedl, R.~Fr\"{u}hwirth\cmsAuthorMark{1}, V.M.~Ghete, N.~H\"{o}rmann, J.~Hrubec, M.~Jeitler\cmsAuthorMark{1}, W.~Kiesenhofer, V.~Kn\"{u}nz, M.~Krammer\cmsAuthorMark{1}, I.~Kr\"{a}tschmer, D.~Liko, I.~Mikulec, M.~Pernicka$^{\textrm{\dag}}$, D.~Rabady\cmsAuthorMark{2}, B.~Rahbaran, C.~Rohringer, H.~Rohringer, R.~Sch\"{o}fbeck, J.~Strauss, A.~Taurok, W.~Waltenberger, C.-E.~Wulz\cmsAuthorMark{1}
% ...
\vskip\cmsinstskip
\dag:~Deceased\\
1:~~Also at Vienna University of Technology, Vienna, Austria\\
2:~~Also at CERN, European Organization for Nuclear Research, Geneva, Switzerland\\
% ...
